# Extreme Software Defined Radio – GHz in Real Time

Eugene Grayver, Alexander Utter
Aerospace Corporation
2310 E. El Segundo Blvd.
El Segundo, CA 90245
310-336-1274
Eugene.Grayver@aero.org

*Abstract*—Software defined radio is a widely accepted paradigm for design of reconfigurable modems. The continuing march of Moore's law makes real-time signal processing on general purpose processors feasible for a large set of waveforms. Data rates in the low Mbps can be processed on low-power ARM processors, and much higher data rates can be supported on large x86 processors. The advantages of all-software development (vs. FPGA/DSP/GPU) are compelling – much wider pool of talent, lower development time and cost, and easier maintenance and porting. However, very high-rate systems (above 100 Mbps) are still firmly in the domain of custom and semi-custom hardware (mostly FPGAs). In this paper we describe an architecture and testbed for an SDR that can be easily scaled to support over 3 GHz of bandwidth and data rate up to 10 Gbps. The paper covers a novel technique to parallelize typically serial algorithms for phase and symbol tracking, followed by a discussion of data distribution for a massively parallel architecture. We provide a brief description of a mixed-signal front end and conclude with measurement results. To the best of the author's knowledge, the system described in this paper is an order of magnitude faster than any prior published result.

### TABLE OF CONTENTS

## 1. INTRODUCTION

Software defined radio (SDR) has gone from a semi-utopian idea [1] to an almost universally accepted technique [1]. The inexorable march of Moore's Law has made it feasible to implement real-time signal processing algorithms entirely in software. One of the major advantages of SDR is the potential to quickly implement new waveforms and algorithms on existing hardware. This advantage is especially valuable for unique and challenging waveforms, and for low volume applications. The enormous investment in time and money required to develop a new ASIC for a cell-phone is justified by the lower achievable power and is amortized over millions of sold items. However, only a few modems are required to support an entire satellite network, making them ideal candidates for SDR.

The three[1] main options for executing the 'software' part of a SDR are [3]:
1. general purpose processors (GPPs),
2. specialized processors such as GPUs,
3. FPGAs.

Going down the list of these choices incurs a significant (up to 10x) increase in NRE. Not only is the development more difficult, but it requires more specialized expertise, making it more challenging to staff. This paper addresses the question:

*What throughput (data rate) is achievable using only GPPs?*

The throughput of a modem is (almost) always limited by the demodulator since it is always more computationally expensive than the modulator. Of course, the answer is a moving target. In fact, the answer has almost doubled since the work on which this paper is based was started.

We set out to achieve data rates at least ten times faster than any previously published result. The target waveform does not address any specific requirement but is meant to be a representative example. The target is a ground station for a high-rate satellite downlink. The hardware must fit into a standard rack, and SWAP is not a major concern. The satellite is sending a continuous waveform based on the DVB-S2 standard [7]. The modulation is 8PSK, coding is rate ½ LDPC/BCH with a block size of 64,800. The symbol rate is 3 GHz, resulting in bandwidth of ~4 GHz, sampled at

---

[1] We do not consider DSPs such as TI TMS320C6657 because of their relatively niche applications – most in cellular base stations.

5 Gsps complex baseband (or equivalently 10 Gsps real IF). The waveform is made up of continuous fixed-length frames, with each frame consisting of a fixed preamble followed by the payload consisting of one FEC block. These requirements can be readily met using a modern FPGA but are rather challenging for a GPP.

No single server is capable of processing this much data. The challenge is therefore to design a system based on mid-range[2] servers that can collectively process the data. The system consists of a data source [4], a network for distributing the data, and a set of servers to process the data. This paper will address each of the components.

## 2. DIVIDE AND CONQUER

Many threads executing across many cores, many processors, and many servers are required to achieve the target throughput. The software architecture is simpler and more efficient if threads execute independently (i.e. do not exchange any data). This is especially important since the threads are split across many servers and inter-server links are relatively slow. The first task is to divide the continuous stream of samples between all the threads. The total number of samples processed by a thread at a time is called a *chunk*. Once a thread gets a chunk of samples, it is busy processing those samples for some time, $T_p$. The processing is described in section 3. The next chunk of samples (including overlap) goes to a free thread. Continuous processing is achieved if the first thread is done by the time the last thread has started processing.

Note that the data source is waveform-agnostic and has no concept of frame boundaries. The FEC decoder requires a complete code block, and therefore a complete frame. Thus, each thread processes an integer number of frames. Since the threads don't initially know where frame boundaries are, there must be some overlap between chunks provided to different threads. The minimum overlap is one frame (i.e. $k$ frames' worth of unaligned samples must be processed to guarantee $k-1$ complete frames). A chunk contains $k$ frames plus a few samples. A few, $K$, extra samples are needed in case the receiver sample rate is slightly faster than 'nominal.' In that case, the timing tracking block drops some samples and generates fewer symbols than expected based on the chunk size and the sample rate. For example, 10 additional samples are needed to compensate for a sample rate offset of $10^{-5}$ (10 ppm), and a chunk size of $10^6$. These extra samples give rise to a few edge conditions: Consider an example with $k=3$ is shown in the Figure 1.

1. A chunk starts on the first sample of a frame
   - chunk 1 includes A,B,C
   - chunk 2 includes D,E
   - chunk 3 includes F,G
2. A chunk just misses the first sample of a frame
   - chunk 1 includes B,C
   - chunk 2 includes D,E
   - chunk 3 includes F,G
3. A chunk starts near the end of a frame
   - chunk 1 includes B,C
   - chunk 2 includes D,E,F
   - chunk 3 includes F,G
4. A chunk starts just before the first sample of a frame. Since the overlap between chunks is slightly larger than a frame:
   - chunk 1 includes A,B,C
   - chunk 2 includes C,D,E
   - chunk 3 includes F,G

Scenarios 1,3,4 show that sometimes a thread ends up an 'extra' frame. The extra frame is not handled by any other thread. There are two issues with handling the extra frame:

- Some threads will take longer to process a chunk of samples than others. This is not a serious problem since all the threads are sharing a set of processors/cores. Note that, as discussed in section 4, we get better performance by locking threads to cores. However, all the cores pull data from a common queue and the extra delay is amortized across all threads.
- The decoder processes 16 frames at a time [5][6]. The

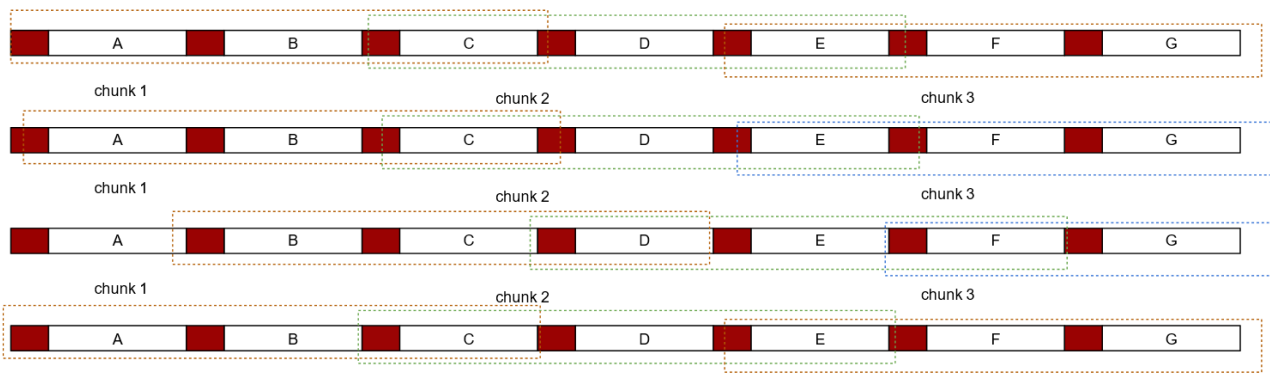

Figure 1. Segmenting input samples into chunks

---

[2] Server price vs. performance increases faster than linear, and there is little benefit in using the highest-end servers since we need more than one.

single extra frame will be padded with 15 'all-zero' codewords. Note that if early-termination is enabled, it is quite likely that the single frame will take less time to decode than a normal 16-frame block because the decoder does not have to worry about 'all-zero' codewords converging.

The extra frame problem is infrequent since it occurs only when the chunk falls in the first $K$ samples of a frame.

The extra frame issue can also be manifested as a duplicate frame. The overlap between chunks is larger than a frame, and it is therefore possible for the same frame to be processed in two chunks. This is also an infrequent event that is handled by the downstream 'stitcher.'

The discussion above has assumed that the chunk size can be any length. However, the numerology is complicated by the constraints of the network subsystem described in section 6.

## 3. SAMPLE PROCESSING

The sample rate processing pipeline follows a classical receiver architecture as shown in Figure 2. Signal processing consists of multiple operations such as phase tracking, timing tracking, etc [9].

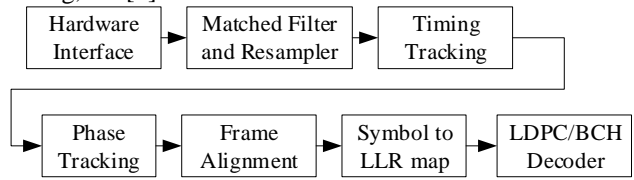

**Figure 2. Receiver Flowgraph**

Multiple chunks of data are processed on a given machine at any time. Two approaches were considered (Figure 3):
a) Threads are dedicated to an operation (e.g. timing tracking). Multiple timing tracking threads get chunks of data as input and generate chunks of data as output.
b) Threads are dedicated to a chunk of data. Each thread does all the operations in sequence.

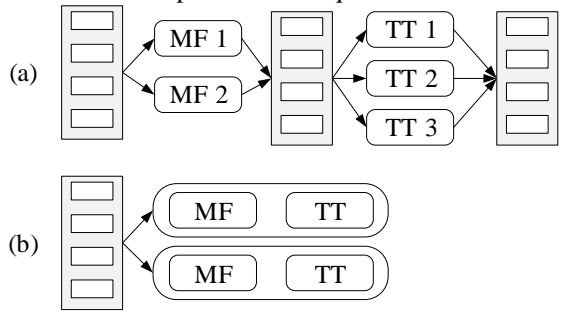

**Figure 3. Two approaches to parallelization**

If the number of chunks processed at a time is close to the number of available cores, the second approach is preferable. One major advantage is that data stays local to a thread, which helps with cache management. The other advantage is that we don't have to worry about the relative throughput rates of different tasks to determine the number of threads allocated to each task. However, the first approach uses less memory since fewer total instances of each block are created. Fortunately, none of the blocks in the demodulator require a lot of memory[3]. The overall latency is the same for both approaches.

Consider an example shown in Figure 4: chunks arrive every time interval; matched filtering (MF) takes 2 time intervals per chunk, and timing tracking (TT) takes 3 time intervals per chunk. Both approaches require 5 threads to keep up with real time.

| Time | 1 | 2 | 3 | 4 | 5 | 6 | 7 | 8 | 9 | 10 | 11 | 12 | 13 | 14 |
|---|---|---|---|---|---|---|---|---|---|---|---|---|---|---|
| Input | a | b | c | d | e | f | g | h | i | j | | | | |
| MF 1 | | a | a | c | c | e | e | g | g | i | i | | | |
| MF 2 | | | b | b | d | d | f | f | h | h | j | j | | |
| TT 1 | | | | a | a | a | d | d | d | g | g | g | | |
| TT 2 | | | | | b | b | b | e | e | e | h | h | h | |
| TT 3 | | | | | | c | c | c | f | f | f | i | i | i |
| | | | | | | | | | | | | | | |
| MF + TT 1 | a | a | a | a | a | f | f | f | f | f | | | | |
| MF + TT 2 | | b | b | b | b | b | g | g | g | g | g | | | |
| MF + TT 3 | | | c | c | c | c | c | h | h | h | h | | | |
| MF + TT 4 | | | | d | d | d | d | d | i | i | i | i | | |
| MF + TT 5 | | | | | e | e | e | e | e | j | j | j | | |

**Figure 4. Timing for two parallelization approaches**

We benchmarked both approaches and selected the second one because it achieves about 20% higher throughput. The next sections provide details about each of the signal processing components.

### Resampling

The data source is waveform agnostic and operates at a constant sample rate. For example, the 4 GSymbol/s waveform is sampled at 5 Gsamples/s resulting in 1.6 samples/symbol. The received samples are first resampled to 2 samples per symbol. An 81-tap resampling filter (5 up, 4 down) also takes care of matched filtering. A rational multirate filter from the IPP library provides an efficient implementation [8]. The resampler also takes care of matched filtering since the coefficients are based on a root-raised-cosine with a cutoff of $\frac{1}{1.6 \times 5} = 0.125$.

### Symbol and Phase Tracking

Adaptive tracking algorithms are inherently serial – the value of the current sample depends on the value of the previous samples. Starting from some initial state, the tracking loop converges to the steady state. The convergence time (i.e. initial transient) is inversely proportional to the loop bandwidth. The loop bandwidth is limited by the loop stability and error variance, which in turn depend on the SNR. The target waveform is designed to operate at SNR around 0 dB, which limits the loop bandwidth to very small values. According to [9], the symbol tracking loop can take up to 100,000 symbols to converge. A simple solution would be to allow the loop to converge and simply discard the samples during the transient. We chose to use a more efficient approach known as multi-pass tracking (Figure 5). Instead of starting at the first sample and moving forward, we start at an offset and move backward. The offset is large enough to guarantee convergence when the algorithm reaches the first sample. The direction of processing is then reversed by changing the sign of the second order term, and the complete chunk is processed. This approach is used for both symbol and phase tracking.

---

[3] The LDPC decoder requires by far the most memory since it processes 16 coded blocks at a time [6].

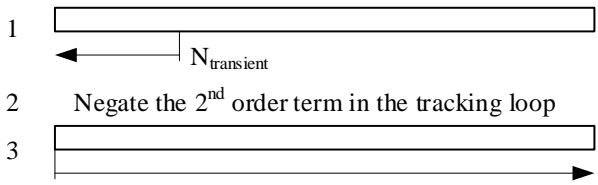

**Figure 5. Two-pass tracking algorithm**

*Symbol Tracking*

A blind (i.e. non-data-aided) symbol tracking loop is used to compensate for the difference between the transmitter and receiver clock rate. A fractional resampler is implemented using an 8-tap Lagrange interpolator. A set of 128 filters is pre-computed to represent equally spaced delays of about 1% of a symbol (Figure 6).

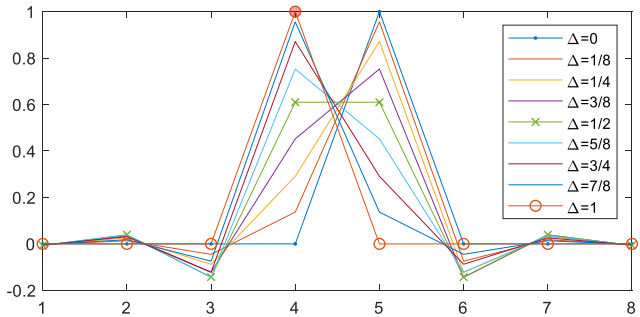

**Figure 6. Tap values for a few of the resampler filters**

The delay (i.e. filter index) is adjusted by the output of the loop filter. A sample is repeated if the filter index wraps around 0 and a sample is skipped if the filter index wraps around 128. It is worth noting that using a single-rate filter or the dot product function from the IPP library is significantly slower than explicitly writing out the 8-tap dot product[4]. The dot product can be efficiently implemented using just a couple of SSE instructions, which are inferred by the compiler (see code below where x_r is the real part of the input and t is the currently selected filter). The overhead of an IPP function call is in fact higher than the entire computation.

```
t = LAGRANGE_ARRAY[current_delay];
for (int i=0; i<num_samples; i++)
    y_r[i]=x_r[0]*t[7] + x_r[1]*t[6] +
           x_r[2]*t[5] + x_r[3]*t[4] +
           x_r[4]*t[3] + x_r[5]*t[2] +
           x_r[6]*t[1] + x_r[7]*t[0];
```

Symbol rate offset is expected to be no more than 10 ppm. This low rate allows us to process *N=64* samples at once and only update the tracking loop once every *N* samples[5]. A FIR filter implementation has an implicit memory (state) equal to the number of taps. Rather than dealing with the memory we chose to process 8 additional samples to effectively 'flush' the memory. Thus, a total of 64+8=72 input values are processed for every 64 outputs. The 12% overhead is more than compensated for by the simplified code. Gardner's timing error detector[6] is computed on the *sum* of *N* early and ontime interpolated values.

*Phase Tracking*

The phase changes much faster than the symbol offset, but it is still reasonable to update the loop once every 8 symbols. This assumption allows us to process 8 samples at once and take advantage of the SSE instructions[7]. The phase for 8 samples is computed by as: $\theta_k = \theta_{-1} + k\varphi$, where $\varphi$ is the frequency. Sine and cosine of the phase are computed using the optimized function from [10]. Note that this function is about 3× faster than using the ippsSinCos_32f_A11 function from IPP. The error is computed using decision directed slicing: $e = \Im(x \times \hat{x}^*)$, where $\hat{x}$ is the estimated symbol. The slicing is implemented using __mm256 intrinsics.

*Frame Synchronization*

Frame boundaries are established by looking for a known fixed preamble. The SNR can be too low to reliably detect an individual preamble. We take advantage of the symbol and phase tracking completed prior to this block and combine the preambles from multiple frames *coherently*. The input chunk is split into frames, arbitrarily starting with the first sample. The frames are then added together to form a vector *x*. The preambles start at an unknown offset, but the offset is the same for each frame. The combined frame is then correlated against the preamble (*p*), and the correlation peak indicates the offset. The correlation is implemented in the FFT domain $(F^{-1}(F(x) \times F(p)^*)$, and the FFT of the preamble is pre-computed.

Samples prior to the first offset and samples after the last complete frame are discarded. The preambles are stripped of the frames and the payloads are passed to the decoder.

---

[4] The excellent performance of the IPP library leads many developers to 'use it by default,' not realizing that a direct implementation may be faster.

[5] Benchmarking showed almost no performance improvement for *N>64*.

[6] $e(n) = y_r(1) * (y_r(2) - y_r(0)) + y_i(1) * (y_i(2) - y_i(0))$ where $y_i$ is the interpolated value.

[7] 8 complex floats can be processed in one operation using the 512-bit AVX instructions.

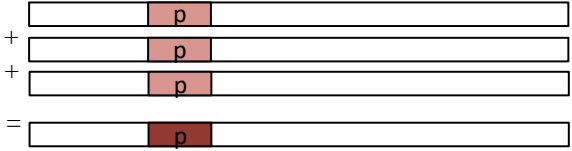

**Figure 7. Combining multiple frames to detect preambles**

*Symbol to Soft Decision*

The received 8-PSK symbols are mapped to three soft decisions (aka log-likelihood-ratios, LLRs). The mapping is implemented using `__mm256` instrinsics. This block is also responsible to deinterleaving the bits according to the DVB-S2 standard. Note that the slicing operation required for phase tracking shares a lot of operations with the soft decision computation. However, since this block is not computationally expensive (see 6) the duplication of work is acceptable.

*FEC Decoder*

The FEC decoder is typically the most computationally expensive block in a non-spread-spectrum demodulator. Fortunately, a highly optimized decoder was developed in [5]. The performance of this decoder on different modern GPPs is reported in [6].

## 4. BENCHMARKING

The throughput was measured on a few modern processors. The benchmark is setup by pre-computing a large set of input data, saving it memory, and providing it to the signal processing chain as fast as it is consumed. The data source thread is executing on a dedicated core. Note that this benchmark does not include the effects of the network interface. As can be seen from Figure 8, one server can achieve almost 500 Msps input rate (which corresponds to 312 Msymbols/s, and about 450 Mbps). Note that the servers are similarly priced, but AMD EPYC provides double the throughput. This result was also observed in [6].

| CPU | CPUs | Cores | Cache (MB) | Freq (GHz) | Year |
|---|---|---|---|---|---|
| E5-2695v4 | 2 | 36 | 45 | 2.1 | 2016 |
| EPYC-7351 | 2 | 32 | 64 | 2.4 | 2018 |

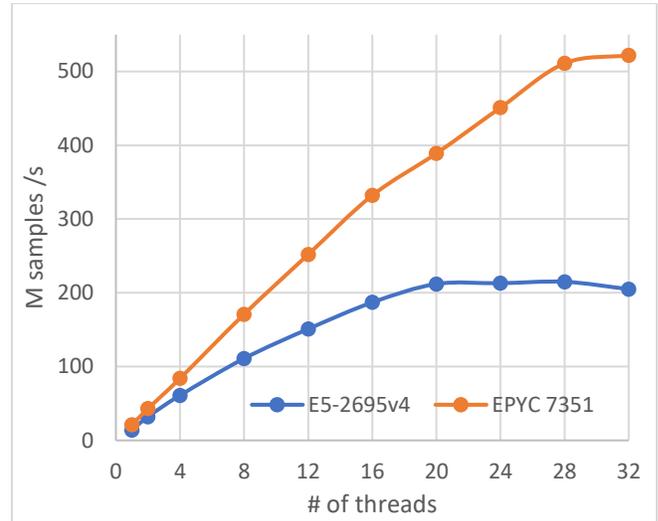

**Figure 8. Throughput for 8-PSK, rate ½ coded system at 1.6 samples/symbol**

The breakdown of CPU utilization by different blocks in the signal processing chain is provided in Figure 9. The decoder takes just over ½ of all the CPU time. Note that according to Ahmdal's law, it would be challenging to significantly increase the throughput since no single block is responsible for the clear majority of the utilization.

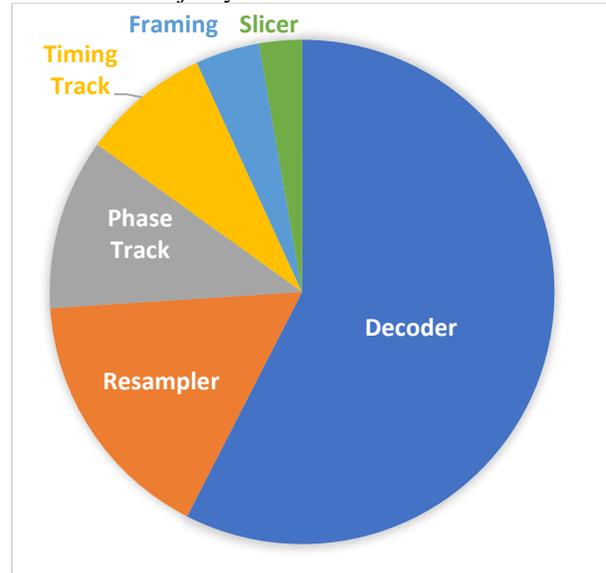

**Figure 9. CPU utilization by block**

The best throughput was achieved by manually assigning threads to cores. Best performance is achieved when 28 of the 32 available cores are used[8,9]. The best assignment resulted in about 8% higher throughput than the worst assignment.

The power consumption scales mostly linearly with the sample rate (Figure 10). Note that these measurements were taken with data coming in over a network interface and therefore account for the power consumption of the network interface card[10].

---

[8] Two cores are used by the operating system, one for input data, one for output date.

[9] Stable (i.e. no dropped packets) operation at sample rates above 300 Msps was achieved only after the signal processing cores were isolated from the kernel by adding `isolcpus=1-15,33-47,17-31,49-64` `nohz_full=1-15,33-47,17-31,49-64` `rcu_nocbs=1-15,33-47,17-31,49-64` to the boot line and setting core affinity to the isolated cores. Further, all non-essential services (timesyncd, anacron, cron, network-manager, ufw, aport, speech-dispatcher, unattended-upgrades, whoopsie) were stopped. Hyperthreading was also disabled.

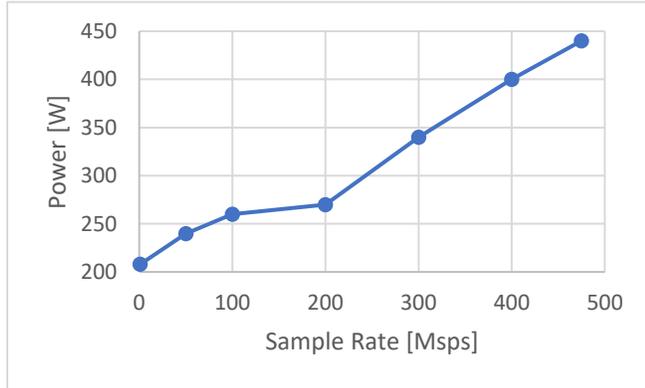

**Figure 10. Power consumption versus sample rate**

## 5. DATA COMBINING

Each server processes data independently of all other servers. At the end of processing a chunk of samples, the server generates 16 or 17 blocks of decoded bits. The decoded bits are sent to the *combiner* server. The combiner server is responsible for putting the decoded bits in order and removing duplicate blocks. The decoded data rate is relatively low (about 450 Mbps). A reliable message passing protocol, ZeroMQ is used to send the decoded bits[11].

The order in which different servers and different threads on a server complete their chunks is not deterministic because the signal processing software executes on a non-real-time Linux[12]. Each block of bits is associated with an absolute *sample number* of the first symbol in the block. The sample number is computed at the start of the sample processing chain based on the absolute packet number. The sample number is then used to determine the order in which decoded bits should be output. The FEC decoder processes 16 blocks at once. However, (see section 2), it sometimes it ends up with 17 code blocks. Let the nominal number of samples per block be *S*. Normally we get 0, 16S, 32S, etc. Sometimes we get 0, 16S, 17S, 33S, etc. Blocks are considered sequential if the delta between their starting sample number is smaller than 17*S*.

Sometimes two code blocks start with the *same* starting sample number. This indicates a repeated frame (see section 2) that must be dropped.

The received blocks are placed into a reordering buffer. The buffer waits for the next sequential block to arrive while buffering out-of-order blocks. However, it is possible for a block to be lost if the SNR is low and the frame synchronization block incorrectly detects frame boundaries. It may also be lost if a server falls behind on receiving UDP packets and an entire chunk of samples is discarded. This condition is handled by setting a maximum number of blocks in the reordering buffer. Once that number is exceeded, the closest *non-sequential* block is output, and processing continues normally.

## 6. NETWORK ARCHITECTURE

The overall architecture is described in [4]. The system (shown in Figure 11) consists of four components:
- **Mixed signal subsystem** converts between baseband or low-IF signal and its digital representation.
- **Real-time signal processing** subsystem converts between the digitized signal and the user data.
- **Recording and playback subsystem** is responsible for recording snapshots of digitized data and/or playing back previously created signals.
- **Network** provides connectivity between the mixed signal and the signal processing and recording/playback subsystems.

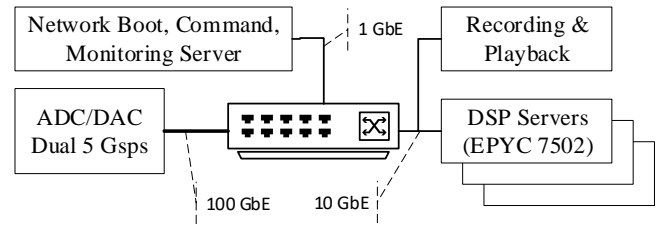

**Figure 11 Overall system block diagram (data plane)**

The input signal is quantized to 8-bit precision, packetized, and sent over a 100 GbE network interface[13]. A switch [15] buffers the high-speed traffic and routes it to multiple servers. Each server has a 10 GbE interface, which is sufficient to receive the maximum achievable sample rate of 500 Msps[14] (see section 4). Two additional network interfaces are used for the control plane and IPMI management. Separating the data from control planes allows us to easily[15] use DPDK [11].

*Numerology*

As described in section 2, the input stream is split between multiple servers with some overlap between servers. The decisions on how data chunks are split into packets, chunk size, and packet size are low-level but important. Each packet has *P* complex samples. Each sample is 2 bytes, making a packet size, *U*, a bit over *2P* bytes. For efficiency

---

[10] The number of active cores was held constant at 28. Additional experiment was run reducing the number of cores to the minimum required to support the sample rate. Power consumption was essentially the same.

[11] Using UDP is definitely feasible, but given the relatively low data rate, ZeroMQ is simpler. The 16 code blocks (~64kB) do not fit into a jumbo frame and would have to be split between multiple packets.

[12] The amount of out-of-order is rather surprising. A thread may complete before another that started dozens of chunks before it.

[13] Total throughput is ~ 5 Gsps × 2 scalars per complex sample × 8 bits per sample = 80 Gbps. Jumbo frames (~8 kB) are used to reduce the overhead due to packet headers.

[14] 8-bit complex samples at 500 Mbps require 8×2×500=8Gbps

[15] A single NIC can be shared between kernel and DPDK using bifurcation but this feature was not used in the testbed.

(and to support AVX-512) packet size should be a multiple of 64 bytes (512 bits).

Multiple packets make up a chunk. The length of each chunk is $N$ samples. The chunk size is a tradeoff between overhead (i.e. larger $N$ means smaller overhead) and CPU cache utilization (i.e. all chunks should fit in cache). The chunk size also has to be smaller than the network switch buffer[16] since the digitizer output rate (100 GbE) is higher than the server input rate (10 GbE). The waveform consists of contiguous frames of length $M$ samples. As discussed in section 2, the overlap between chunks must be just over $M$ samples.

We want to use large packets to reduce the number of packets per second. There is no compelling reason to make $U$ an integer multiple of the frame size. However, arithmetic is simpler if U is close to an integer multiple. Let $U=64\times136=8704$. There are about 8 packets per frame ($8\times8704=34816\times2$). As discussed in 2, the FEC decoder processes 16 frames at once for efficiency. Thus, each chunk must be at least $16M$ samples.

```
Sample rate: 1.6 samples per symbol
Modulation: 8PSK --> 3 bits per symbol
Frame size: DVB-S2 --> 64800 bits = 21600
symbols
Header size: 90 symbols
Frame + Header size: 21690 symbols
Frame + Header size ~  34704 samples
Frames/chunk: 17

Samples/packet: 4352
Packets/frame ~ 7.97 = 8
Packets/multicast group = 8
Samples/multicast group = 1 frame + 112 samples

Packets/chunk = 8*17 = 136
Samples/chunk = 8 frames + 896 samples
```

The packets are sent as UDP multicast. The multicast lets the switch take care of sending overlap segments to two servers at the same time[17]. We send just over one frame (8 packets) per MC group. Unicast approach would require about 6% more network throughput and would also make the digitizer implementation slightly more complex since some packets have to be sent to two destinations while others only to one destination.

Let us consider an example with two servers, S1 and S2:
- S1 subscribes to groups: 0-16
- S2 subscribes to groups 0, 16-31

S1 waits until the first packet in group 0 and then starts grabbing samples. S2 waits until the first packet in group 16 and then starts grabbing samples. In general, a system with $S$ servers will use 16$S$ multicast groups (400 for 25 servers, the switch can support 5000 groups).

## 7. DEBUGGING AND OBSERVATION

A debug and monitoring interface is extremely valuable for any communications system. Diagnosing problems is much simpler if an engineer has access to real-time monitoring points as well as the ability to record snapshots for post processing. The recording capability is discussed in [4]. The GUI (Figure 12) provides a tree view of the available monitor points sorted by either name/host/thread or host/thread/name.

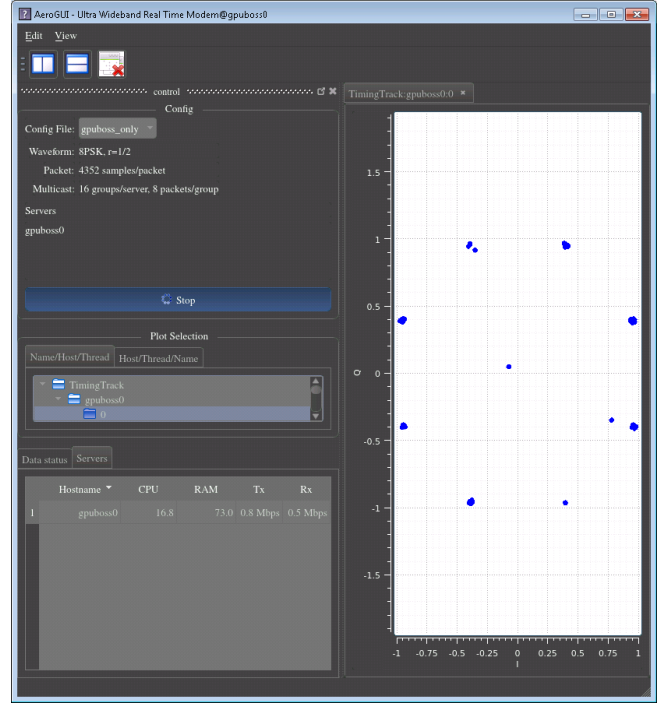

**Figure 12. Monitoring and debugging GUI**

The real-time monitoring is somewhat complicated because it needs to support dozens of threads executing on dozens of servers. It may be necessary to simultaneously look at different points in the signal processing chain on different servers. The monitoring interface should consume no CPU resources until it is used and should consume minimal CPU when used. A monitor block is inserted at the output of every signal processing block (a total of #of threads × #of signal processing blocks). Each block creates a ZeroMQ publish socket and connects to the monitoring server[18]. It then sends out periodic (e.g. 1s) message advertising the existence of this monitor and its parameters: name, type, thread id, host id, address for the monitor socket. The monitor server (GUI) connects to the advertised address using the ZeroMQ request/reply protocol. Once a request is received, the monitor block captures samples into a buffer (buffer size is set in the request message) and returns the captured buffer. This approach ensures that the monitor is blocked waiting for a request and not using any CPU.

---

[16] Our switch has 16 MB of buffer space shared between all the ports, but each port is limited to 2 MB of buffer.

[17] Modern NICs implement multiple hardware queues, each handled by a separate interrupt, for load balancing. The default rule for assigning packets to a queue uses the source and destination IP addresses. A multicast group is defined by an IP address, and different groups get routed to different queues. This is highly undesirable because the packets arrive out-of-order (if using the UDP network stack) or must be retrieved using separate calls if using DPDK. The default queue assignment rule must be changed to force all of our packets into a single queue. This is done using the `ethtool -U $nic flow-type udp4 dst-port 4660 action 1` command for UDP network stack or using the DPDK flow IP.

[18] We use a PUB socket because don't want packets to be queued or blocked.

## 8. CONCLUSION

The testbed described in this paper demonstrates that it is feasible to implement a real-time SDR capable of processing multiple GHz/Gbps on standard servers running unmodified Linux. This implementation is very well suited for a satellite ground station. The same servers can be used to support a wide range of downlinks and can be time shared between multiple downlinks. Technology refresh and upgrades are dramatically simplified since the same software will run on newer hardware without even recompiling. As processors become ever faster (or have more cores), the number of required servers goes down.

## BIOGRAPHY

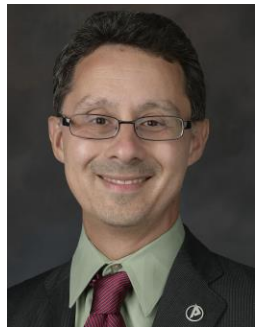

*Dr. Eugene Grayver* received a B.S. degree in electrical engineering from Caltech, and a Ph.D. degree from UCLA in 2000. He was one of the founding team members of a fabless semiconductor company working on low-power ASICs for multi-antenna 3G mobile receivers. In 2003 he joined The Aerospace Corporation, where he is currently working on flexible communications platforms. His research interests include reconfigurable implementations of digital signal processing algorithms, adaptive computing, and system design of wireless data communication systems. He is also participating in the software-defined radio community, trying to define a common configuration standard and determine optimal partitioning between software and hardware. In 2012 he published a book called "Implementing Software Defined Radio."

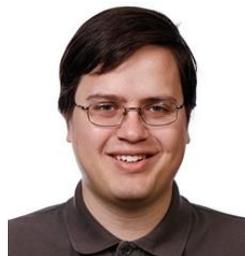

*Alexander Utter* holds a B.S.E. from Harvey Mudd College and an M.S.E.E. from Stanford University. He joined the Aerospace Corporation's Digital Communication Implementation Department (DCID) in 2007. His research interests include full-stack design and development of low-SWaP embedded space systems, as well as real-time software defined receivers for radio-frequency and free-space optical communications.